# Bibliometric Profile of Nursing Research in Ex Yugoslavian Countries

Helena Blažun Vošner[1,2,3], Peter Kokol[4], Danica Železnik[2], Jernej Završnik[1,5,6]

[1] Zdravstveni dom dr. Adolfa Drolca Maribor, Ulica talcev 9, 2000 Maribor
[2] Fakulteta za zdravstvene in socialne vede Slovenj Gradec, Glavni trg 1, 2380 Slovenj Gradec
[3] Alma Mater Europaea, Slovenska ulica 17, 2000 Maribor
[4] Univerza v Mariboru, Fakulteta za elektrotehniko, računalništvo in informatiko, koroška cesta 46, 2000 Maribor
[5] Univerza v Mariboru, Fakulteta za naravoslovje in matematiko, Koroška cesta 160, 2000 Maribor
[6] Znanstveno raziskovalno središče Koper, Garibaldijeva ulica 1, 6000 Koper

***Abstract***

The development of modern nursing and consequently nursing research in Ex- Yugoslavia is about a century old. To profile the development, volume and content of nursing research in this are we completed a performance and spatial bibliometric analysis combined with synthetic content analysis to identify the most productive countries and institutions, most prolific source titles, country cooperation, publication production trends, content of research and hot topics. The corpus was harvested from the Web of Science – All databases and contained 1380 papers. Slovenia was the most productive country, followed by Croatia and Serbia. The synthetic content analysis demonstrated that nursing research in ex-Yugoslavian countries is growing both in scope and number of publications, notwithstanding the fact that research content differs between countries and it seems that each country is focused on their local health problems. Substantial part of the research is published in national journals in national languages however, it is noteworthy to note that some ex-Yugoslavian authors have succeeded in publishing their research in top nursing journals. The study also revealed substantial international co-operation especially among ex-Yugoslavian countries and European Union.

**BACKGROUND**

The development of modern nursing and consequently nursing research in Ex- Yugoslavia is about a century old. The more intensive, however still slow development started in 1921 when first two schools in nursing were formed in Belgrade and Zagreb. More rapid development started after the Second world war with the raise of the quality and scope of the health services standards. In individual republics the more exhaustive development in nursing as a profession and nursing research started in:

- 1899 in Serbia when the first school for midwives was founded at the Department of Gynecology and Obstetrics of the General State Hospital in Belgrade. However, there were no other schools for nurses in Serbia until the foundation of the School for Midwives of the Red Cross Society in 1921 (Vlaisavljević et al., 2014)
- 1921 Croatia with the founding of the School for Nurse Assistants, however the roots go back to 1882 when Dr. Manšek published a book Voluntary Military Health Care Service (Kalauz et al., 2012)
- 1927 in Slovenia with the establishmet of the Nursing School Graduate Organization at the Institute for Maternal and Child Health (Zbornica zdravstvene in babiške nege Slovenije, 2021). However the establishment of the first midwifery school in Ljubljana in 1753 on the initiative of the Austrian Empress Maria Theresa dates even eralier and presents a big step also in the professionalization of the nursing profession is Slovenia.

Single journals (Kokol et al., 2017) or country groups or wider region based bibliometric research profile studies are recently gaining in popularity . Onyancha and Onyango (O.b & Onyango, 2020) presented the employment of information and communication technologies for agricultural activities in Sub Saharan Countries, Bambo and Pouris (Bambo & Pouris, 2020) investigated the research profile of bioeconomy in South Africa, Lin et al (Gege Lin et al., 2018) analysed the research preferences in G20 countries, Živković and Panić (Živković & Panić, 2020) analysed the development of science and education in Balkan countries and Naruetharadhol and Gebsombut (Naruetharadhol & Gebsombut, 2020) examined the research on food tourism performed in Southeast Asia.

Similar studies, however concerned with medical and health sciences were done for Caribbean and Latin American countries focusing on stroke (Alarcon-Ruiz et al., 2019) and COVID – 19 (Espinosa et al., 2020), Arab Region regarding mental health (Zeinoun et al., 2020) and COVID-19 (Zyoud, 2021) , Southeastern Europe focusing on Liver Transplatation (Mrzljak et al., 2020), Armenia concerned with anaemia in children and adolescent (Awe et al., 2021) and Asia Pacific region concerned with palliative care (Cheong et al., 2018). While there were also some similar studies focusing on nursing research (Benton & Brenton, 2020; Dardas et al., 2019; Thelwall & Mas-Bleda, 2020; Zhang et al., 2018), no study has covered the ex-Yugoslavia countries region, thus the aim of this paper is to cover this gap.

In that manner we performed a performance and spatial bibliometric analysis, and synthetic content analysis (Kokol et al., 2021) to identify the most productive countries and institutions, most prolific source titles, country cooperation, publication production trends, content and structure of research and hot topics.

## METHODS

Bibliometrics is defined as the quantitative analysis of the bibliographic features of a body of literature like journals, monographs, reports, theses, conference papers and similar (Hawkins, 2001; Pritchard, 1969; Železnik et al., 2017). One of its main advantages is the domain independence, and it has been already successfully used in nursing and healthcare (Kokol et al., 2020; Kokol & Blažun Vošner, 2019).

**Search strategy and data analyses**

The various corpora of publications employed in this study were retrieved from the Web of Science (WoS) bibliographical database (Elsevier, Netherlands) on 6th of June 2021. We used two bibliographic collections, namely the Core Collection (WoS CC) and All Databases (WoS AD). WoS AD covers more source titles, but it lacks some of the bibliometric attributes provided by WoS CC, like data for corresponding/first authors analysis or advanced bibliometric mapping. For both collections the search was performed by the Advance search command *TS = Nursing*, thus limiting the search to nursing topic papers. No additional search limitations were set. The full WoS CC based search corpus was used for the content analysis of the publications from the whole region. The corpus was then further partitioned into seven country corpora, one for each ex-Yugoslavian country. These corpora were used for the county specific research content analysis. The next corpus was formed by removing all publications where the corresponding/first authors affiliation was geographically not located in an ex - Yugoslavian country. This corpus was used to analyse and compare the publication bibliometric features per country. like average number of authors, citations, references or pages per paper. The WoS AD corpus was used for descriptive, spatial and thematic bibliometric analysis for the whole ex-Yugoslavia region.

For the descriptive and spatial bibliometric analysis, we used WoS Refine function and MS Excel (Microsoft, United States of America (USA)). For the Excel based analysis publications meta-data was exported to Excel program, where trends' analyses of publication characteristics were performed. Synthetic content analysis using the triangulation of bibliometric mapping (based on author keywords) and thematic analysis (Blažun Vošner et al., 2020; Kokol et al., 2021) and country co-operation was performed with the help of VOSviewer software version 1.6.15 (Leiden University, Netherlands) (van Eck & Waltman, 2010). Hot topics were identified using the approach proposed by Kokol et al (Kokol et al., 2018)

## RESULTS

The search in WoS AD resulted in 1380 publications. Among them there were 1127 articles, 119 review papers, 21 editorials, 18 letters, 15 case reports, 15 clinical trials eight books and 308 other types of publications (please note that WoS can  categorise a single publication in more than one category). The number of retrieved publications for of ex Yugoslavian countries is shown in Table 1. The search in WoS CC resulted in1068 publications – 686 of those have corresponding or first authors from ex- Yugoslavian countries.

First nursing paper published in ex Yugoslavian countries indexed in WoS was published in 1991 by Croatian affiliated authors. The paper presented a study about chromosomal abnormalities among nurses occupationally exposed to antineoplastic drugs (Milkovickraus & Horvat, 1991). After that the number of publications started to grow slowly till 2004, when the trend remained linear. but the growth rate become steeper. In 2020 we can note a slight decrease in productivity. According to the ratio between articles, reviews and conference papers and the linear trend of the literature we reason that nursing research is in the second of four stage of the Schneider scientific discipline evolution model (Shneider, 2009).  That means that the country specific nursing research terminology and methodologies are starting to be standardised, and that domain specific original knowledge generation is heading toward more intensive research productivity. The positive research productivity trend is the largest in Slovenia, and lowest in Kosovo and Montenegro.

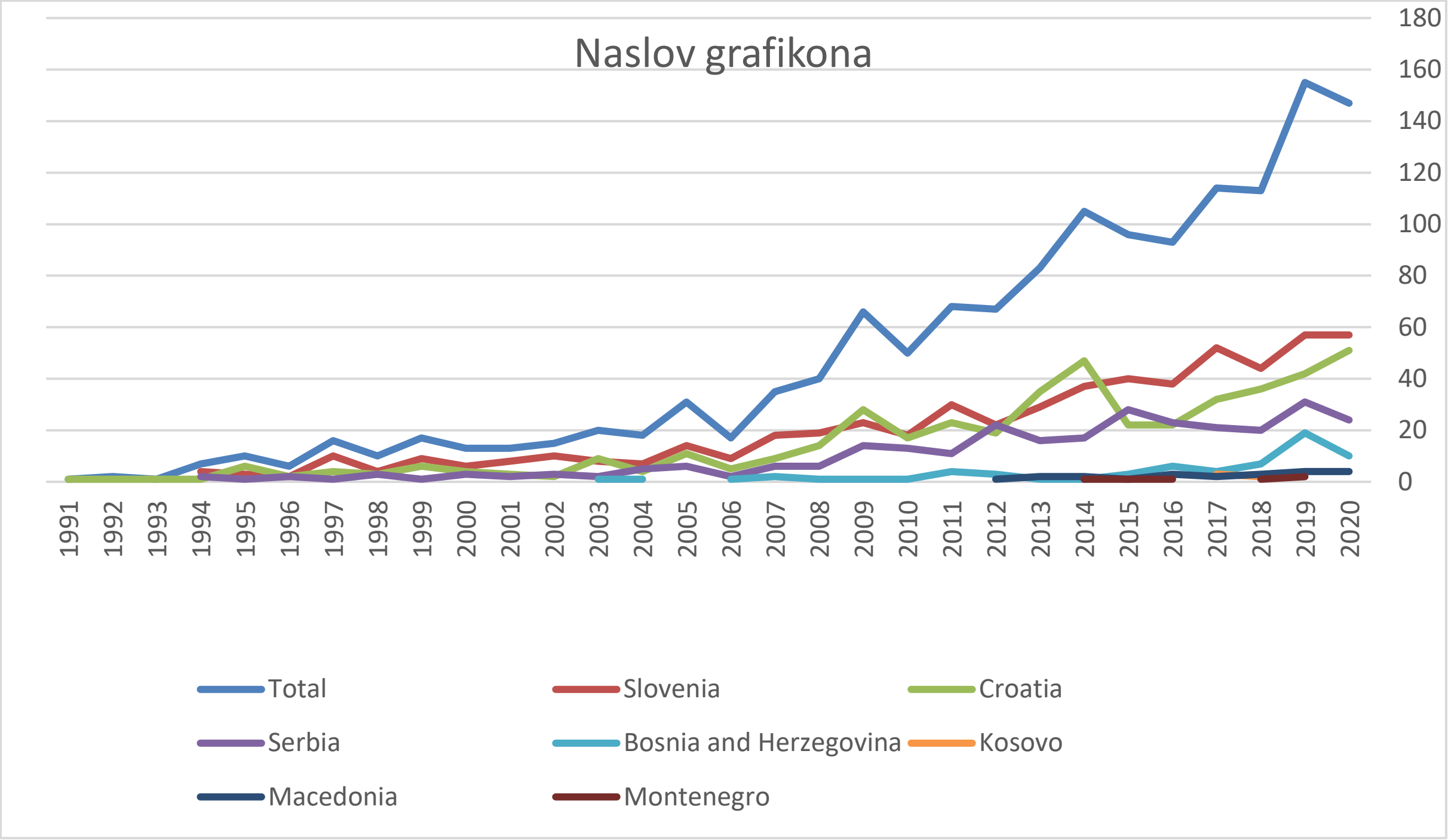

Figure 1: The trends in scientific literature production in ex YU-countries

Bibliometric characteristics of publications where corresponding or first authors are affiliated to an ex-Yugoslavia country are shown in the Table 1. The average number of authors per paper ranges from 3.33 (Slovenia) to 4.82 (Bosnia & Hercegovina). The average number of references per paper ranges from 25.25 (Macedonia) to 76 (Kosovo). The average number of citations per paper ranges from 0 (Kosovo) to 19.49 (Serbia). The average number of references per paper ranges from 25.25 (Macedonia) to 76 (Kosovo). In average the longest papers are published by Montenegro authors (n=14.67) and the shortest by Macedonian authors (n=5.75).

Table 1. Bibliometric characteristics of publications where corresponding or first authors are affiliated to an ex-Yugoslavian country

| | Number of publications | Authors | References | Citations | Pages |
|---|---|---|---|---|---|
| **Slovenia** | 304 | 3.33 | 32.53 | 5.13 | 8.49 |
| **Serbia** | 131 | 4.82 | 35.63 | 19.49 | 10.24 |
| **Croatia.** | 226 | 4.46 | 31.27 | 6.48 | 8.79 |
| **Bosnia & Hercegovina** | 18 | 4.72 | 25.78 | 3.06 | 7.89 |
| **Montenegro** | 3 | 3.67 | 35.67 | 1.33 | 14.67 |
| **Kosovo** | 1 | 4 | 76 | 0 | 13 |
| **Macedonia** | 5 | 3.75 | 25.25 | 0.8 | 5.75 |

**Spatial distribution**

The geographical distribution of nursing research literature production of the countries from the region of ex-Yugoslavia is shown in the Table 2. Slovenia was the most productive country, followed by Croatia and Serbia. Those three countries share of papers where the corresponding or first authors are affiliated to an institution from an ex-Yugoslavian country exceeds 60%. The literature production in the above three countries was substantially larger than in the rest of the region.

Table 2. Geographical distribution of nursing research literature production in ex-Yugoslavian countries

| Ex-Yugoslavia Country | Number of Publications indexed in WoS AD corpus | Number of Publications indexed in WoS CC corpus | Number of Publications in WoS CC corpus where corresponding/first author is from ex-Yugoslavia |
|---|---|---|---|
| **Slovenia** | 600 | 472 | 304 |
| **Croatia** | 476 | 380 | 226 |
| **Serbia** | 303 | 227 | 131 |
| **Bosnia and Hercegovina** | 71 | 55 | 18 |
| **Macedonia** | 30 | 26 | 3 |
| **Kosovo** | 7 | 1 | 1 |
| **Montenegro** | 7 | 5 | 3 |

Top ten most productive institutions were University of Ljubljana. University of Zagreb and University of Maribor. Most productive institutions are located in Slovenia (n=4), Croatia (n=4) and Serbia (n=2).

Table 3: Most productive institutions in ex - YU-countries

| Institutions | Publications |
|---|---|
| **University of Ljubljana** | 342 |
| **University of Zagreb** | 302 |
| **University of Maribor** | 251 |
| **University of Belgrade** | 189 |
| **University medical centre Ljubljana** | 90 |
| **University of Primorska** | 72 |
| **University of JJ Strosmayer Osijek** | 68 |
| **University of Rijeka** | 67 |
| **University Spli** | 62 |
| **University of Novi Sad** | 48 |

Most prolific source titles regarding the whole ex-Yugoslavian territory publishing 20 or more publications were: *Acta Medica Croatica* (n=62), *Zdravstveno varstvo* (34), *Studies in Health Technology and Informatics* (n=30), *Nurse Education Today* (n=28), *Collegium Antropolgicum* (n=22), *European Journal of Cancer* (n=20) and *HealthMed* (n=20). Among those four are local professional journals, publishing papers also in national languages. Most prolific country specific source titles are listed in the Table 4. The majority of the most prolific journals are national journals from different health and medical sciences fields. The most prolific international source titles are *Studies in Health Technology and Informatics* and *Nurse Education Today*.

The most prolific nursing journals per individual country are presented in Table 4. And Table 5. Slovenia, Croatia and Serbia were the most active in publishing in nursing professional journals. In addition to Nurse Education Today, Nursing Ethics and International Nursing Review were the most prolific journals.

Table 4: Most prolific source titles

| Croatia | Bosnia | Slovenia | Serbia | Macedonia | Montenegro | Kosovo |
|---|---|---|---|---|---|---|
| **ACTA MEDICA CROATIA (52)** | LECTURE NOTES IN NETWORKS AND SYSTEMS (4) | ZDRAVNISKI VESTNIK (36) | VOJNOSANITETSKI PREGLED (15) | JOURNAL OF ADVANCED NURSING (2) | ACTA MEDICO HISTORICA ADRIATICA (1) | CLINICAL INTERVENTIONS IN AGING (1) |
| **COLLEGIUM ANTROPOLOGICUM (21)** | BREAST (2) | ZDRAVSTVENO VARSTVO (29) | SRPSKI ARHIV ZA CELOKUPNO LEKARSTVO (13) | BMJ OPEN (1) | ARCHIVES OF INDUSTRIAL HYGIENE AND TOXICOLOGY (1) | CONFLICT AND HEALTH (1) |
| **ACTA CLINICA CROATICA (16)** | HEALTHMED (2) | NURSE EDUCATION TODAY (23) | HEALTHMED (12) | BRITISH JOURNAL OF HEALTH PSYCHOLOGY (1) | NEUROPSYCHIATRIC DISEASE AND TREATMENT (1) | DEMENTIA AND GERIATRIC COGNITIVE DISORDERS (1) |
| **ARHIV ZA HIGIJENU RADA I TOKSIKOLOGIJU (13)** | INTERNATIONAL NURSING REVIEW (2) | STUDIES IN HEALTH TECHNOLOGY AND INFORMATICS (21) | EJC SUPPLEMENTS (15) | EUROPEAN STROKE JOURNAL (1) | ONCOLOGIST (1) | INTERNATIONAL JOURNAL OF NURSING PRACTICE (1) |
| **CROATIAN MEDICAL JOURNAL (13)** | JOURNAL OF ADVANCED NURSING (2) | EUROPEAN JOURNAL OF CANCER CARE (17) | NURSE EDUCATION TODAY (5) | JOURNAL OF SLEEP RESEARCH (1) | POZNAN STUDIES IN CONTEMPORARY LINGUISTICS (1) | JOURNAL OF ALZHEIMER S DISEASE (1) |

Table 5: Most prolific nursing specific source tiles

| Total | Slovenia | Croatia | Serbia | Bosna and Hercegovina | Macedonija | Kosovo |
|---|---|---|---|---|---|---|
| **Nurse Education Today (28)** | NURSE EDUCATION TODAY (23) | NURSE EDUCATION TODAY (6) | NURSE EDUCATION TODAY (5) | INTERNATIONAL NURSING REVIEW (2) | JOURNAL OF ADVANCED NURSING (2) | INTERNATIONAL JOURNAL OF NURSING PRACTICE (1) |
| **Nursing Ethics (13)** | NURSING ETHICS (9) | JOURNAL OF ADVANCED NURSING (4) | INTERNATIONAL NURSING REVIEW (2) | JOURNAL OF ADVANCED NURSING | NURSE EDUCATION IN PRACTICE (1) | |
| **International Nursing Review (11)** | JOURNAL OF CLINICAL NURSING (9 | EUROPEAN JOURNAL OF CARDIOVASCULAR NURSING(3) | PAIN MANAGEMENT NURSING (4) | NURSING ETHICS | NURSING ETHICS (1) | |
| **Journal of Advanced Nursing(10)** | INTERNATIONAL NURSING REVIEW (7) | INTERNATIONAL NURSING REVIEW (3) | EUROPEAN JOURNAL OF CARDIOVASCULAR NURSING (2) | INTERNATIONAL JOURNAL OF NURSING STUDIES (1) | | |
| **Journal of Clinical Nursing (10)** | JOURNAL OF NURSING MANAGEMENT (7) | NURSE EDUCATION IN PRACTICE (3) | INTENSIVE AND CRITICAL CARE NURSING (2) | NURSE EDUCATION IN PRACTICE (1), NURSING EDUCATION PERSPECTIVES (1) | | |

**Content analysis**

Thematic analysis revealed that nursing research in Ex – Yu countries is concerned with following themes (Figure 2) Occupational exposure of health workers (dark blue colour), Education of health professionals (green colour), Nursing education, research and informatics (light blue colour), Job satisfaction of health personal (yellow colour), Quality of life (orange colour), Care for elderly (Scarlett colour) and Quality of healthcare performed by multidisciplinary teams (Red colour). Historically (Figure 3), the research before 2012 was concerned with studies on occupational exposure of health workers and nursing informatics, in the next period with medical education, public health, depression, elderly, ethics, and patient satisfaction and safety, then with nursing education, job satisfaction, quality of health care and quality of life and in the recently with bibliometrics, mental health, long term care, competencies and quality indicators.

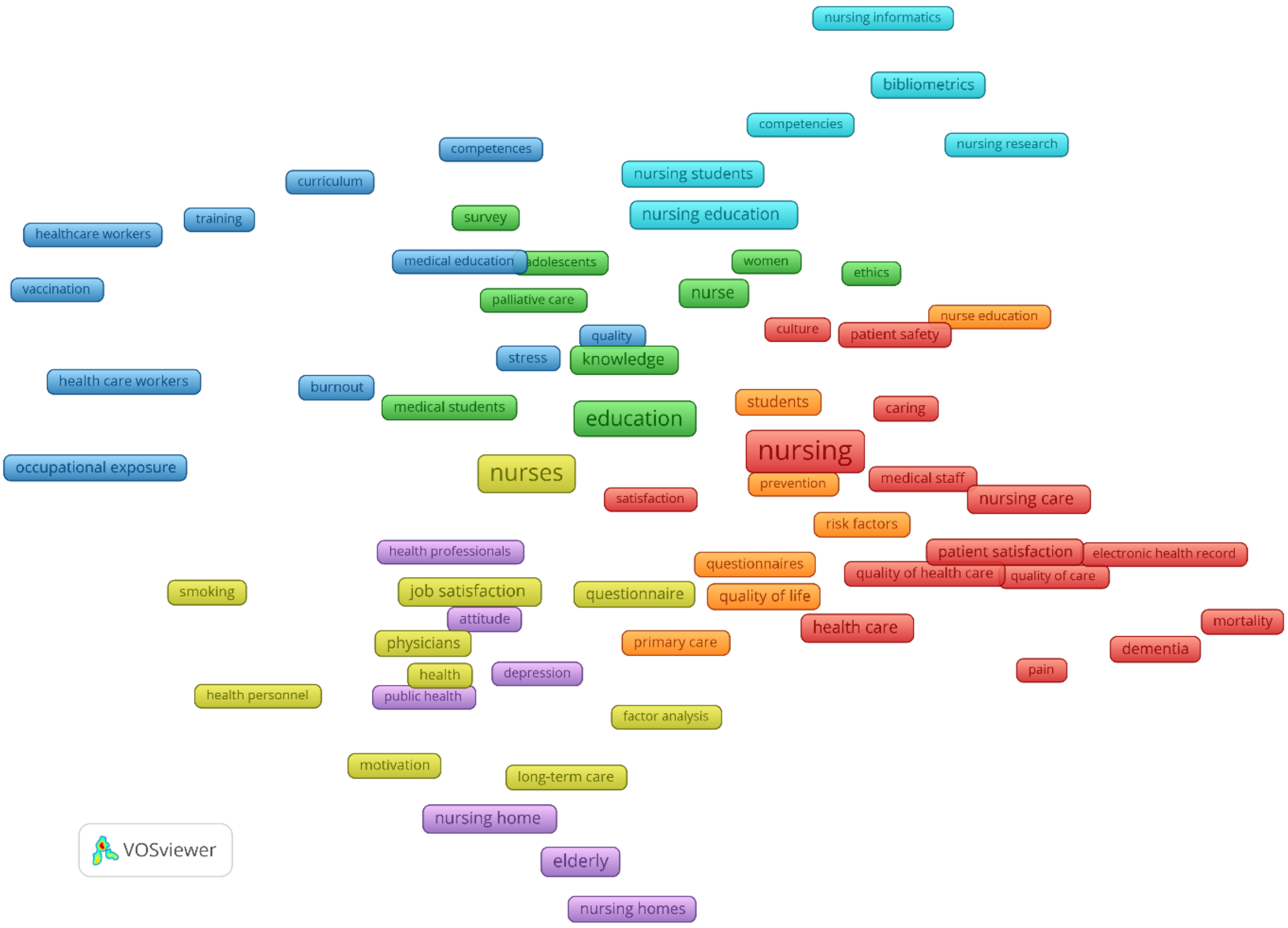

Figure 2. The authors keywords cluster landscape

The early research (before 2012) in the ex-Yugoslavia region (Figure 3) was concerned with occupational exposure/satisfaction of health workers, curriculum development and nursing informatics and education. Between 2012 and 2016 the research was focused on similar themes, however new research themes, like care for elderly, vaccination, palliative care, primary health, quality of life and quality of care become popular. In the recent period some hot topic emerged. Most popular are bibliometrics, pain management and long term care.

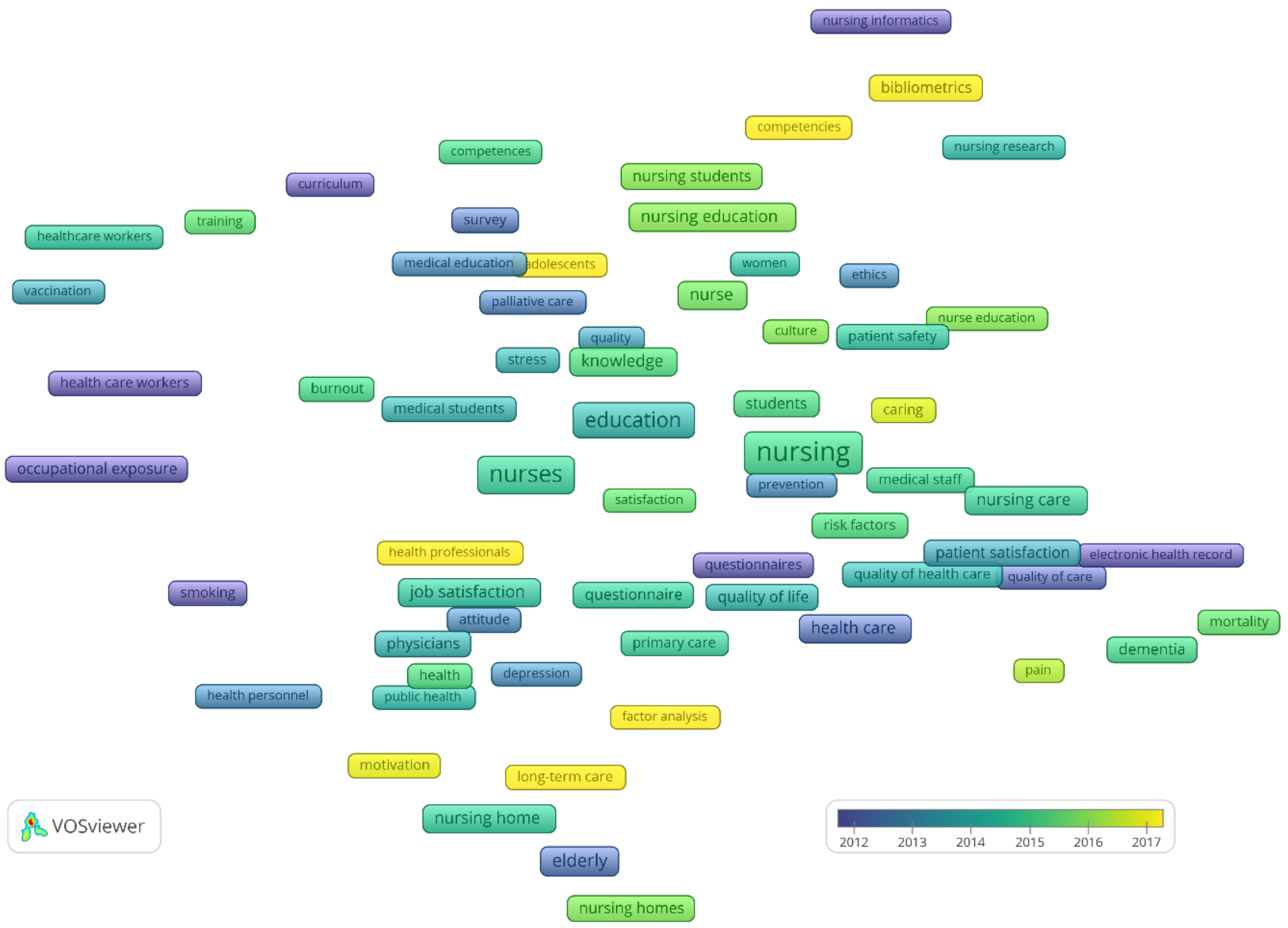

Figure 3. The timeline of the nursing knowledge development in the Ex-Yugoslavia region

The differences in the content and themes of nursing research between ex-Yugoslavian countries is shown in Table 6. The analysis of the table reveals that each country has its own country specific research focus.

Table 6. Thematic and content comparison of nursing research in ex-Yugoslavian countries

| Country | Slovenia | Croatia | Serbia | Bosna and Hercegovina | Macedonia | Kosovo | Montenegro |
|---|---|---|---|---|---|---|---|
| Main research themes | Bibliometrics, Nursing care for elderly, community nursing, e- | Occupational exposure, cardiovascular diseases, Nursing education, Quality | Elderly care, patient education and satisfaction, Job satisfaction, | Nursing interventions for refuges and migrants, Missed and Unfinish | Ethics of nursing care, Missed and unfinished nursing care, Knowledge integration | Absenteeism of nurses | Oncology |
| | learning in nursing education, patient safety and satisfaction, community nursing | of healthcare | Interprofessional relations, Occupational exposure of healthcare workers | ed nursing care | in multidisciplinary content | | |
| Most popular keywords | Nursing education, Nursing home, Bibliometrics | Education, Attitudes, prevention | Education, health care workers, Job satisfaction | Refuges, missed nursing care, nursing interventions | Unfinished nursing care, rationing | Sickness | Cancer, Health care budget, incidence |
| **Most cited keywords** | Burnout, citation Analysis, Attitude | Guidelines, Stress, Work ability | Cardiac arrest, Mortality, Nursing home | Missed nursing care, Education | Prioritisation, Ethics | N/A | N/A |
| **Hot topics** | Citation analysis, long term care, evidence based practice | Older adults, unfinished nursing care, mental health | Dementia, Case based reasoning, Nursing home | Unfinished nursing care | Unfinished, missed and undone nursing care | N/A | Oncology |

**Country co-operation**

The ex Yugoslavian countries were internationally extremely active (Fig 4.) – they established a cooperation co-authorship network of 88 countries. Among ex-Yu countries. most co-authorship links were instituted by Serbia (n=66) followed by Croatia (n=63), Slovenia (n=61), North Macedonia (n=55), Bosna and Herzegovina (n=33), Montenegro (n=16) and. Kosovo (1). The strongest links between ex Yu countries were established between Slovenia and Croatia (n=32), and the strongest link between ex-Yu country and non ex Yu. Country between Croatia and England (n=40).

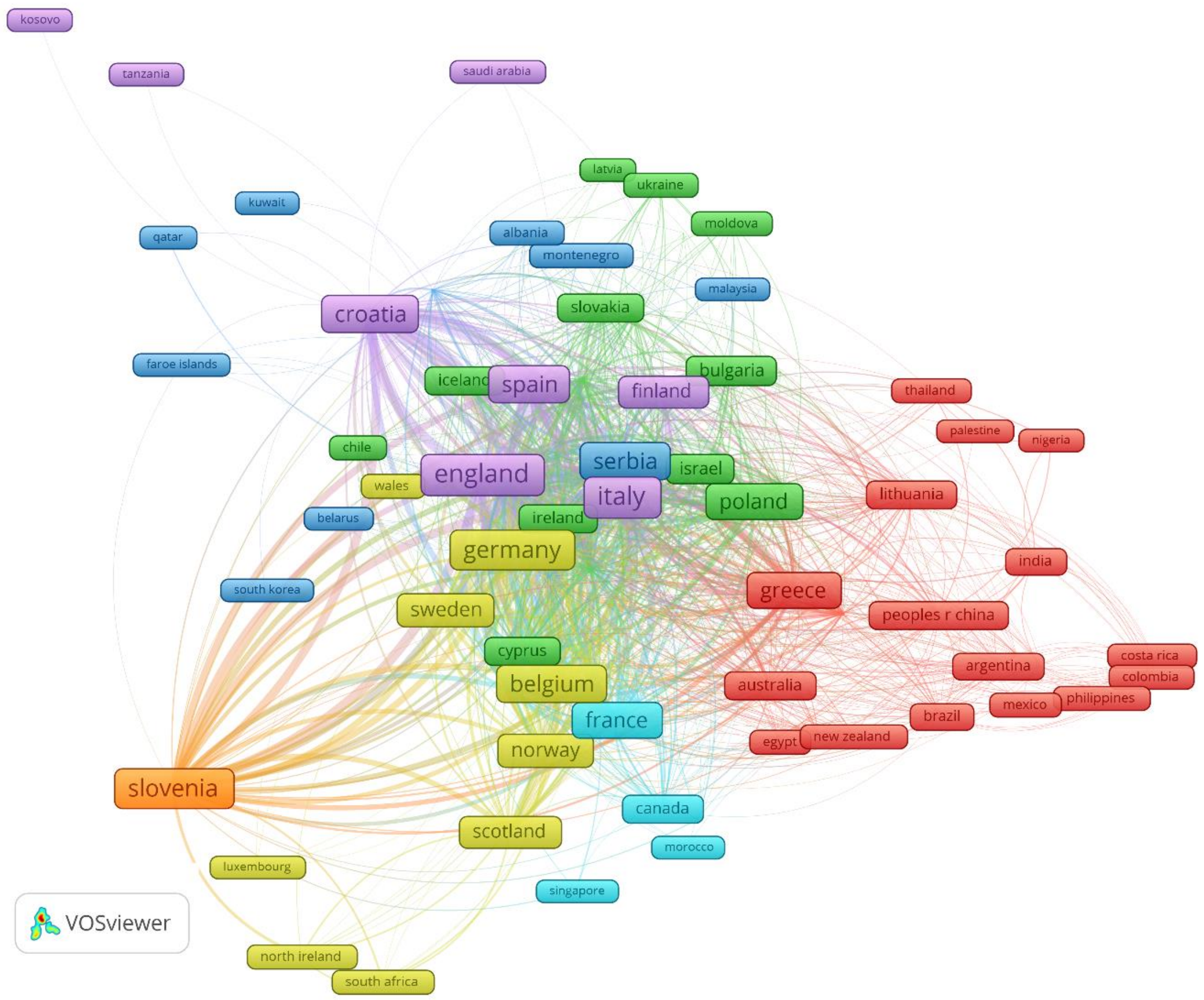

Figure 4. The international collaboration of Ex Yugoslavian countries

**DISCUSSION**

Performance and production of individual ex Yugoslavian countries broadly correspond to the overall scientific production of those countries where Croatia is on 50th rank, Serbia 52th, Slovenia 53 th, Bosnia and Herzegovina 96th, North Macedonia 97th and Montenegro 122th. It is also interestingly to note that the countries which established nursing schools early in the 20th century are significantly more scientifically productive than the countries which established them after the second world. This seems to be quite logical as nursing research is an important part of the .educational process, the process which also empowers nurses to perform the research (Tingen et al., 2009). The ex Yugoslavian countries with higher economic determinants/indices are considerably more productive. This correspondence with outcomes of Kokols et al study (Kokol et al., 2019) about the relation between country and health determinants and nursing research productivity and also with the study of the research productivity of Arab countries regarding COVID-19 research (Zyoud, 2021).

Substantial part of the research is published in national journals in national languages; however it is noteworthy to note that some ex-Yugoslavian authors have succeeded in publishing their research in top nursing journals like International Journal of Nursing Studies, Journal of Advanced Nursing, Nursing Education Today and Nursing Ethics. Strong international cooperation which emerged in the region, whilst strongest between ex Yugoslavian countries themselves, but notable cooperation also exist with some EU countries. The strong international co-operation ight help researchers to extend their success in publishing in top journals in the future. Comparing nursing to areas such as medicine, molecular biology and genetics, where new discoveries are made daily, nursing advances at a slower rate. Consequently, publishing in  more general and extremely high impact journals like Nature, Science, The Lancet, JAMA, or New England Journal of Medicine is yet to come. To enable such publishing endeavours, not only international but interdisciplinary research must emerge in the region. Accordingly, policy makers in national and international funding agencies and bodies should put more focus on including nursing in interdisciplinary research grants calls. Additionally, the preparation of high quality nursing research publications requires significant effort and time, however it is worth doing because  publishing in high-impact journals enables researchers to generate further funding and thus support collaborative research needed to achieve above goals and subsequently and most important improve region health determinants (Grant & Buxton, 2018). Our study also showed that **m**ost of the research is performed at universities and clinical centres. Thus, to further alleviate above points the nursing research in the region should spread to all types of health care institutions.

Our analysis demonstrated that ex Yugoslavian countries nursing research is growing both in scope and number of publications, notwithstanding the fact that research content differs between countries and that it seems that each country is focused on their local health problems. Given the increasingly global nature of nursing research, these findings highlight another need for region and EU policy makers and funding bodies, namely to allocate research funding in such manner that it will support the inclusion of ex-Yugoslavia countries nursing research into the broader scientific community. This will enable the translation of global nursing knowledge to the region and also enable region researchers to contribute to the global nursing knowledge development.

On the other hand we shall not forget about the political, economic and cultural differences between ex-Yugoslavia countries, and that nursing research in the region should contribute to culturally appropriate nursing care that will improve health and well-being through the whole region. That means that for example Islamic patients should receive nursing care according to Islamic principles, beliefs, and values, not only in their home countries, but at least in their neighbouring countries, further emphasising the need for knowledge transfer between ex Yugoslavian countries and worldwide.

**Study limitations and strengths**

Our study did have some limitations. First, the use of other bibliographic databases like Scopus might lead to different results than, due to the fact that different databases uses different categorisations and standards foe country and institution names, and also differ in the list of information titles covered Additionally, the synthetic content analysis was performed on information source abstracts and titles only; it is therefore possible that the results could have been different if the whole publications would be available for analysis. The synthetic content analysis enables the minimisation of the interpretative bias, however some bias is still possible and may affected the outcomes of the content analysis. On the other hand, the holistic bibliometric analysis has never been performed for the region of ex-Yugoslavian countries. The study also revealed several characteristics and specifics of ex-Yugoslavian countries nursing research, and enabled us to identify similarities and differences between countries, which is another strength of our study.

**CONCLUSION**

Our study revealed some interesting facts about the nursing research in the ex Yugoslavia region, which might enable researchers, academics, clinical nurses, policy makers, funding agencies and government leaders to enhance efficiency of future studies and understand further applications of nursing research for improving health determinants and wellbeing. Understanding the nursing knowledge development in the region and differences between countries might be beneficial not just from the scientific point of view, but also to evidence-based policy-making. To improve the nursing research, researchers in the region especially from low and middle income countries must develop collaborations with researchers in high-income countries and also focus to integrate nursing research into interdisciplinary .knowledge development endeavours. Collaborative and interdisciplinary research not only contributes to a more comprehensive understanding of nursing science problems, but might contribute to building a shared base of data, innovations, evidence and research paradigms. Finally, we must state that more research funds should be dedicated to conducting research on this important topic.